\documentclass[twocolumn,prX,nofootinbib]{revtex4-2}
\pdfoutput=1
\usepackage{amsmath,amssymb,amsfonts,eucal,mathrsfs,graphicx,tensor}
\usepackage[usenames,dvipsnames]{color}
\usepackage{hyperref}
\usepackage{subfigure}
\DeclareMathAlphabet{\mathpzc}{OT1}{pzc}{m}{it}

\definecolor{grey}{rgb}{0.75,0.75,0.75}

\definecolor{brown}{rgb}{0.65,0.16,0.16}

\newcommand{\beql}[1]{\begin{equation}\label{#1}}
\newcommand{\eeq}{\end{equation}}

\newcommand{\dr}{\mathrm{d}r}
\newcommand{\ds}{\mathrm{d}s}
\newcommand{\dt}{\mathrm{d}t}
\newcommand{\dT}{\mathrm{d}T }
\newcommand{\dv}{\mathrm{d}v}
\newcommand{\dy}{\mathrm{d}y}
\newcommand{\dx}{\mathrm{d}x}

\newcommand{\lag}{\mathscr{L}}

\newcommand{\uhor}{\text{\sc uh}}
\newcommand{\ruh}{r_{\uhor}}

\newcommand{\HLg}{Ho\v{r}ava-Lifshitz gravity}
\newcommand{\uhbh}{universal horizon black hole}

\begin{document}
\title{Energy-entropy relation for asymptotically Lifshitz spacetimes with universal horizons}
\author{Jonathan Cheyne}\email{jacheyne85@googlemail.com}
\author{David Mattingly}\email{david.mattingly@unh.edu}
\affiliation{Department of Physics and Astronomy, University of New Hampshire, Durham, NH 03824, USA}
%*****************************************************************************************************************************************
\begin{abstract}
	We numerically solve for 2+1 asymptotically Lifshitz universal horizon solutions in \HLg~for dynamical exponents $z=2$ through $z=8$.  We find that for all $z$ there is a thermodynamical first law and Smarr formula.  Furthermore, we find that the energy-entropy relation expected for a thermal state in a two dimensional Lifshitz field theory, $E=\frac{2}{z+2}TS$, is also satisfied for universal horizons, including the correct $z$ scaling. 
\end{abstract}

\maketitle

\section{Introduction}

If holography is to provide as much calculational utility as possible, it is necessary to find duals to field theories which are not subject to conformal symmetry, as such field theories occur in numerous systems~(examples in \cite{Julian:1996,Stewart:2001zz,Si:2008,Kachru:2008yh,Si:2011hh}). One minimal deviation from conformal invariance, which therefore has potentially valuable applications, is Lifshitz symmetry.

The appropriate gravitational dual to a Lifshitz field theory must admit solutions which somewhere in the bulk obey Lifshitz symmetry. Since these type of solutions cannot be found in vacuum general relativity, this requires modification on the bulk side of the duality. We can generate such solutions with a particular choice of fields and profile in the bulk, for example in Einstein-Dilaton-Maxwell theory~\cite{Goldstein:2009cv} or with a Proca field~\cite{Taylor:2008tg} but these types of relativistic duals for Lifshitz theories are less preferred for multiple reasons.  They are asymptotically AdS and so correspond to a conformal UV completion,  run counter to the notion that zero temperature field theories are dual to vacuum spacetimes, and fail to naturally provide duals for entanglement entropy~\cite{Gentle:2015cfp,Cheyne:2017bis}.

Alternatively, one can generate both globally and asymptotically Lifshitz spacetimes as solutions to an appropriately modified theory of gravity. Lifshitz solutions have been found in massive gravity~\cite{AyonBeato:2009nh} and bi-gravity theories,~\cite{Goya:2014eya} but in these models the Lifshitz nature is a feature of the solutions and not built-in to the theory itself at a fundamental level.  If the gravitational theory is naturally Lifshitz in the UV and admits asymptotically Lifshitz solutions, this would resolve issues such as those with matching of Weyl anomalies~\cite{Horava:2009vy}, and with entanglement entropy~\cite{Cheyne:2017bis}. \HLg~is just such a theory. It requires a preferred foliation of spacetime in order to include the necessary higher spatial derivative terms in the action that make it UV Lifshitz while remaining ghost free, but the absence of ghosts and renormalizability of the theory provides a putative complete field theoretical model of quantum gravity. Furthermore it generically admits asymptotically Lifshitz solutions with a tuneable parameter, usually called the Lifshitz exponent $z$, which can be taken as equivalent to the corresponding dynamic exponent on the field theory side of the duality.   

The introduction of finite temperature on the field theory side of a duality is associated with the introduction of a black hole to the gravitational bulk. Since it has been shown that there exist asymptotically Lifshitz black hole solutions to \HLg \cite{Basu:2016vyz}, one might reasonably expect these to correspond to finite temperature Lifshitz field theories on the boundary. Consequently the thermodynamics of such field theories ought to be directly related to that of the black holes in the bulk.

In this paper we derive two primary results. First, we show that 2+1 dimensional, asymptotically Lifshitz universal horizon black hole solutions to \HLg~for $z=2$ through $z=8$ exist (and of course for higher $z$ as well in all probability).  Second, and more importantly, the numerical coefficient in the energy-entropy relation varies with respect to the dynamical exponent, $z$, in exactly the manner one expects if the boundary dual is a 2-d Lifshitz field theory. Specifically, the general thermodynamic relation
\begin{equation}
	E=\frac{d}{z+d}TS,\label{eqn:Peet}
\end{equation}
where $E$ is the energy density, $T$ the temperature, and $S$ the entropy density, and $d$ is the dimension, holds for $d=2$. This relation has previously been shown \cite{Bertoldi:2009dt} to also hold for black holes in spacetimes generated with Einstein-Maxwell-dilaton theory.  However, as mentioned above, since relativistic duals to Lifshitz field theory fail for other dynamic observables, but \HLg~duals do not, our result can be understood as confirmatory evidence that  \HLg~may provide the appropriate dual for generic $z$ Lifshitz theories.

The structure of the paper is as follows.  We briefly review \HLg~in section \ref{sec:HLg}, discuss the existence of global and asymptotically Lifshitz \uhbh~solutions and describe the numerical procedure in section \ref{sec:priors}, outline our findings for various $z$, the corresponding first laws, and the energy-entropy relation in section \ref{sec:results}, and conclude in section \ref{sec:implications}.  Throughout this paper we use metric signature $(-,+,+)$, Greek indices will refer to $d+1$ dimensional spacetime quantities, while Latin indices will refer to $d$ dimensional spatial quantities. 

\section{Ho\v{r}ava-Lifshitz gravity and equations of motion}\label{sec:HLg}
 \HLg~is a proposed theory of quantum gravity~\cite{Horava:2008ih,Horava:2009uw} (for a review see~\cite{Sotiriou:2010wn}) which is power counting renormalisable and ghost free. It achieves this through the addition of higher spatial derivative terms in the propagator, which implies that it lacks Lorentz invariance in the UV limit.  The differing treatment of spatial and temporal derivatives in the action necessarily requires a clear decomposition of the spacetime into spatial and temporal directions at any point, and thus the foliation of the spacetime into leaves of simultaneity, with each leaf labelled by some monotonic scalar function $T$.
 
 The theory is required to be invariant under monotonic transformations $\tilde{T}(T)$ which preserve this foliation, and can be made generally covariant and well-behaved (the so-called \textit{healthy extension}~\cite{Blas:2009qj}) through the promotion of $T$ to a dynamical field (the khronon) in the action and corresponding incorporation of all possible kinetic terms. Since the theory must be invariant under monotonic transformations of $T$, $T$ cannot appear directly in the action.  Instead, what appears in the action is the unit one form $u_a$, the \textit{\ae ther}, everywhere normal to the surfaces of constant $T$, and defined by
\begin{equation}
u_\mu=-N\nabla_\mu T,
\end{equation}
where the lapse $N$ is given by
\begin{equation}
    N=(-\nabla_\mu T\,\nabla^\mu T)^{-1/2}
\end{equation}

Given the aether, the action for the covariant healthy extension of Ho\v{r}ava-Lifshitz gravity in a $d+1$ decomposition is
  \begin{eqnarray} \label{eqn:hlg}
 	S=\frac{1}{16\pi G}\int\dT\int_{\Sigma_T}\mathrm{d}^{d}\tilde{x}N\sqrt{\tilde{g}}\bigg(\tilde{R}-\Lambda+\alpha a^2\nonumber\\+\beta K^{ab}K_{ab}-\gamma K^2-V(\tilde{g}_{ab},N)\bigg)
 \end{eqnarray}
 where $\tilde{R}$ is the $d$-dimensional Ricci scalar, $K_{\mu \nu}$ is the extrinsic curvature of the hypersurfaces, $K_{\mu \nu}=\nabla_\mu u_\nu +u_\mu a_\nu$, $a_\mu$ is the acceleration of the aether, $a_\mu=u^\nu \nabla_\nu u_\mu$, and the tilde denotes evaluation purely over the spatial directions. $V(\tilde{g}_{ab},N)$ contains the set of higher dimension operators built out of fields and spatial derivatives $D_a=(g_a^b+u_a u^b) \nabla_b$.
 
 Since we are concerned with $1+2$ dimensional black hole solutions with curvatures everywhere much smaller than any high energy Lifshitz scale, it will suffice to consider the infrared limit of $1+2$ dimensional \HLg~and to restrict the action to terms quadratic in spatial derivatives, dropping $V(\tilde{g}_{ab},N)$.
 With this restriction, and as a result of the Ho\v{r}ava-Lifshitz action being definable in terms of the aether, the infrared truncated theory is closely related~\cite{Jacobson:2010mx} to Einstein-aether theory~\cite{Jacobson:2000xp}, which is the most general two derivative theory for a unit timelike vector field coupled to gravity. 
 
 To keep notation consistent with the literature~\cite{Basu:2016vyz,Berglund:2012bu,Berglund:2012fk,Pacilio:2017emh,Liberati:2017vse,Bhattacharyya:2014kta}, we will formulate the infrared truncated healthy extension of \HLg~in the notation of Einstein-aether theory.  
  The full action of Einstein-aether theory is,

 \begin{equation}
 S=\frac{1}{16\pi G}\int\mathrm{d}^{d+1}x\sqrt{-g}\left(R-2\Lambda+\lag_{\ae}+\lambda(u^2+1) \right).
\end{equation} 
 where
 \begin{equation}
 \lag_{\ae}=-Z^{\mu\nu}{}_{\sigma\rho}(\nabla_\mu u^\sigma)(\nabla_\nu u^\rho),
 \end{equation}
\begin{equation}
 Z^{\mu\nu}{}_{\sigma\rho}=c_1g^{\mu\nu}g_{\sigma\rho}+c_2\delta^\mu{}_\sigma\delta^\nu{}_\rho+c_3\delta^\mu{}_\rho\delta^\nu{}_\sigma-c_4u^\mu u^\nu g_{\sigma\rho}.
 \end{equation}
 and  $\lambda$ is a Lagrange multiplier used to enforce the normalisation condition $u^2=-1$.
 
 The aether in \HLg~is automatically normalized and hypersurface orthogonal as a consequence of its relation to $T$.  Hence one overall combination of kinetic terms (that of the twist) and the Lagrange multiplier make no contribution to the equations of motion.  \HLg~is then equivalent to the reduced action of ``T-theory''~\cite{Jacobson:2010mx}, which in Einstein-aether notation is simply the Ho\v{r}ava-Lifshitz action (\ref{eqn:hlg}) with
 \begin{eqnarray}
 	\alpha&=&c_{14}=c_1+c_4
 \\	\beta&=&1-c_{13}=1-(c_1+c_3)
 \\	\gamma&=&1+c_2.
 \end{eqnarray}

 Given this equivalence, the equations of motion can be read off from Einstein-aether theory (c.f. the derivation in \cite{Pacilio:2017emh}) as
\begin{subequations}
\begin{align}
&\nabla_\mu(N\underleftarrow{\AE^\mu})=0 \\
\begin{split}
&G_{\mu \nu}+\Lambda g_{\mu \nu}=c_1(\nabla_\mu u_\gamma\nabla_\nu  u^\gamma-\nabla_\gamma u_\mu \nabla^\gamma u_\nu)+c_4a_\mu a_\nu\\
&+\nabla_\gamma X\indices{^\gamma_{\mu \nu}}-(u\cdot \AE) u_\mu u_\nu -2\AE_{(\mu}u_{\nu)}+\frac{1}{2}L_u\,g_{\mu \nu},
\end{split}\label{eq:st:eom:2}
\end{align}
\end{subequations}
Here the underleft arrow means to project onto the hypersurfaces of constant $T$ and the notation has been shortened by use of the quantities
\begin{subequations}
\label{eq:ae:yx}
\begin{align}
&\AE_\mu=c_4\,a^\gamma\nabla_\mu u_\gamma+\nabla_\gamma Y\indices{^\gamma_\mu},\\
&Y\indices{^\mu_\nu}=Z\indices{^{\mu\gamma}_{\nu \delta}}\nabla_\gamma u^\delta,\\
&X\indices{^\gamma_{\mu \nu}}=u^\gamma Y_{(\mu \nu)}+u_{(\mu}Y\indices{^\gamma_{\nu)}}-u_{(\nu}Y\indices{_{\mu)}^\gamma},
\end{align}
\end{subequations}

Due to the hypersurface orthogonal nature of the aether vector, each of the terms above will upon evaluation become proportional to $c_{14}, c_{13}$ or $c_2$.  We now turn to the relevant solutions for \HLg.

\section{Lifshitz solutions and their generation}\label{sec:priors}
\subsection{Global Lifshitz}
The Lifshitz spacetime with line element
\begin{equation}
	\ds^2=-\left(\frac{r}{l}\right)^{2z}\dt^2+\left(\frac{r}{l}\right)^2\dx^2+\left(\frac{l}{r}\right)^2\dr^2
\end{equation}
and aether vector parallel to $d/dt$ everywhere is a vacuum solution to the equations of motion and captures the spacetime symmetries of Lifshitz type QFTs with dynamical exponent $z$ in 1+1 dimensions \cite{Griffin:2012qx}.  $l$, the Lifshitz length scale, and $z$, the Lifshitz exponent, are related to the parameters in the action by
\begin{eqnarray}
	l^2=-\frac{z(z+1)}{2 \Lambda}\\
	z=\frac{1}{1-c_{14}}
\end{eqnarray}

It is preferable that the global Lifshitz solution naturally exist as a vacuum spacetime. If we wish the bulk state corresponding to a boundary state in thermal equilibrium, i.e. a black hole solution, to be smoothly connected to its zero temperature equivalent, then it is necessary for the desired asymptotics to exist as part of a vacuum solution on the gravitational side. We note that most other methods of forming Lifshitz spacetimes rely on a non-zero matter field profile. 
%While there are other issues with these solutions discussed below, it is worth considering the possibility that if the fields used to generate these solutions couple to other content in the action, they may not be stable against perturbations, and thus may relax over time into states with lower energy content, rendering them unsuitable as duals to the zero temperature boundary theory.
Other evidence for \HLg~as the natural (finite temperature) dual includes the matching of the anisotropic Weyl anomalies through holographic renormalisation of the theory~\cite{Griffin:2012qx}. Furthermore, holographic calculation of entanglement entropy shows that the appropriate closure of entanglement wedges in global Lifshitz spacetime can be achieved using \HLg~\cite{Cheyne:2017bis}, where attempts with relativistic theories fail~\cite{Gentle:2015cfp}.

\subsection{Asymptotically Lifshitz universal horizon spacetimes}

As the UV Lifshitz nature of \HLg~implies arbitrarily fast high energy modes in the aether frame, one might expect black hole solutions to be impossible. However, there exist solutions that do indeed possess trapping horizons for even these ultra high energy modes~\cite{Blas:2011ni}, and some of these solutions are asymptotically Lifshitz\footnote{In fact, if a cosmological constant is present Lifshitz solutions are the \textit{generic} solutions with planar symmetry as asymptotically AdS solutions exist only for the special case $c_{14}=0$.~\cite{Bhattacharyya:2014kta}}~\cite{Basu:2016vyz}.  These  solutions are static with a timelike Killing field $\chi$ and either planar or spherical symmetry. At some radius in the bulk the solution is such that $u \cdot \chi=0$, which implies that the Killing vector is tangent to a constant $T$ hypersurface.  Since even ultrahigh energy modes propagate forward in $T$, at this radius a trapping surface occurs where all excitations propagate towards smaller r.  This surface is called the \textit{universal horizon} and can be shown to obey a variant of the first law of black hole thermodynamics in both the asymptotically flat and Lifshitz cases, radiate thermally, etc.   

For our purposes the asymptotically Lifshitz solutions to \HLg~are most relevant.  Spacetimes with $z=2$ asymptotics and planar universal horizons have been found numerically, with corresponding evidence that there exist the usual law of black hole thermodynamics~\cite{Basu:2016vyz}.  Our goal is to extend the prior numerical $z=2$ work to higher $z$.

\subsection{Procedure for numerically generating solutions}
We obtain $z\geq 2$  asymptotically Lifshitz spacetimes by following a procedure very much like that in \cite{Basu:2016vyz}, which we briefly review here. We first assume planar symmetry, and write the metric in Eddington-Finkelstein style in terms of functions of the radial coordinate, $r$,
\begin{equation}
	\ds^2=-e(r)\dv^2+2f(r)\dv\dr+r^2\dy^2
\end{equation}
The Killing vector $\chi^\mu=\partial_v$ is the time translation Killing vector associated with staticity, and $\partial_y$ with translational symmetry in the transverse coordinate, $y$.  The metric has two degrees of freedom, $e(r)$ and $f(r)$, to be solved for.  

The aether itself has one degree of freedom under this symmetry, that of its ``tilt'' in the radial direction.  In order to move from the covariant equations of motion to the coordinate representation and capture this degree of freedom it is useful to introduce the unit one-form $s_\mu$, orthogonal to $u$ and the transverse direction, such that $s^\mu$ points in the direction of radial infinity. We can then relate $\chi, u_\mu$, and $s_\mu$ via the coordinate specific functions 
\begin{eqnarray}
	u_\mu=(u\cdot\chi)\dv+\frac{f(r)}{(u\cdot\chi)-(s\cdot\chi)}dr\\
	e(r)=(u\cdot\chi)^2-(s\cdot\chi)^2\\
	X(r)=(s\cdot\chi)-(u\cdot\chi)
\end{eqnarray} 
where the top equation is the decomposition of the aether one-form in these coordinates, the second equation comes from the unit norm constraint, and the last equation is a choice of a third function $X(r)$ that captures the aether tilt in an algebraically convenient manner.  Given these relations and the three functions $e(r), f(r), X(r)$ one can explicitly rewrite the covariant action and equations of motion in terms of them, as in equation (12) of~\cite{Basu:2016vyz}.  Furthermore, by use of $X(r)$ as defined above, the equation of motion for $f(r)$ becomes algebraic, so $f(r)$ can be solved for substituted back into the equations for $e(r)$ and $X(r)$. This leaves two second order differential equations for $e''(r)$, and $X''(r)$ that must be numerically solved.  The form of the equations is long and not very enlightening, so we forgo them here.

These two equations have singularities at a particular value of $r$, at which there exists a trapping surface for the additional spin zero mode which propagates in \HLg. This spin zero mode propagates along geodesics of an effective metric $g^{(0)}_{\mu\nu}=g_{\mu\nu}-(s_0^2-1)u_\mu u_\nu$, where $s_0^2$ is the square of the speed of the spin zero mode. Therefore a requirement that $s_0^2=1$ will colocate the spin zero and Killing horizons. Since the action is invariant under a disformal redefinition of the field~\cite{Foster:2005ec}, which only modifies the values of the $c_i$ coefficients, $s_0^2$ can be adjusted without changing the underlying solution space (other solutions, with $s_0^2\neq1$, can then be found afterwards by performing the inverse disformal transform). We will implicitly make this redefinition and set $s_0^2=1$ to simplify calculations. 

Doing so fixes the coefficient $c_{13}$ via
\begin{equation}
s_0^2=1=\frac{4z}{(1-c_{13})(n_s(z-1)-2(z-1))(n_s(z+1)-4)}
\end{equation}
where $z$ is the dynamical exponent, and $n_s$ is an integer which encodes the order at which $s\cdot\chi$ for the spacetime diverges from that for global Lifshitz \cite{Basu:2016vyz}. Additionally, asymptotically Lifshitz behavior with analytic falloffs is also a solution to the \HLg~field equations only if $c_2$ satisfies the constraint~\cite{Basu:2016vyz}
\begin{equation}
	\frac{c_{13}+c_2}{1-c_{13}}=\frac{4(z-1)}{n_s(n_s-2)(z+1)^2}
\end{equation}
and $c_{14}$ satisfies
\begin{equation}
	c_{14}=\frac{z-1}{z}.
\end{equation}
In summary, having set $l=1$ and $s_0^2=1$, the choice of $z$ and $n_s$ then fixes the coefficients $c_2$, $c_{13}$, and $c_{14}$ in terms of $z$ and the integer $n_s$.

Regularity at the spin-0 horizon reduces the parameter space further beyond just fixing the $c_i$ coefficients.  Requiring that $e''(r)$ and $X''(r)$ are regular at $r_0$, the spin zero horizon radius once $s_0^2=1$ has been set, establishes a further equation that relates $e(r_0)$, $e'(r_0)$, $X(r_0)$, $X'(r_0)$ to each other and to $r_0$, $z$, and $n_s$. In addition to this constraint we have the requirement that $e(r_0)=0$, since the spin zero horizon coincides with the Killing horizon, and freedom to set $X(r_0)$ to an arbitrary value as this corresponds to an overall normalization of $\chi$ that we can fix after the solution is found.  Therefore the naive four dimensional initial value parameter space is actually one dimensional. In practice, we choose $X(r_0)$ and $n_s=4$, and then for each combination of $r_0$ and $z$ of interest, we numerically search the parameter space of $X'(r_0)$ imposing regular spin zero horizons to find the unique spacetime which asymptotically approaches global Lifshitz.   

In order to search this parameter space we employ an iterative procedure, essentially the shooting method with initial data at the spin zero horizon $r_0$. We begin by expanding the expressions for $e(r)$ and $X(r)$ around $r_0$ out to fourth order in $(r-r_0)$.  By substituting each coefficient into the next order terms we are able to express all higher derivatives in terms of $X(r_0)$, $X'(r_0)$, $z$, $r_0$, and $n_s$. With these, we are able to choose $X(r_0)$ and evaluate $e(r_0+\delta r)$ and $X(r_0+\delta r)$ (where $\delta r=10^{-5}r_0$) analytically to generate initial data slightly away from the singular point of the equations. We then evolve $e$ and $X$ numerically outwards, until the evolution breaks down (at some $r_{break}>r_0$). We repeat this for various values of $\frac{X'(r_0)}{X(r_0)}$ until we find the value for which $r_{break}>10^3r_0$. This is a spacetime which, to an extremely high degree of precision, is asymptotically Lifshitz. At this point, we use the same procedure to evolve the solution inwards to the radial location ($\ruh$) of the universal horizon (the outermost point at which $u\cdot\chi=0$).

We then perform an overall normalisation on the values of $e$ and $X$, which is done by requiring that asymptotically
\begin{eqnarray}
	\lim_{r\to\infty}\frac{e(r)}{r^{2z}}=1
\\	\lim_{r\to\infty}\frac{f(r)}{r^{z-1}}=1
\\	\lim_{r\to\infty}\frac{u\cdot\chi}{r^{z}}=-1.
\end{eqnarray}
The above procedure generates the unique asymptotically Lifshitz universal horizon spacetime for some choice of $r_0$. We then vary $r_0$ to generate a family of solutions with different values of the universal horizon radius.  These characterize the complete behavior of the family of asymptotically Lifshitz universal horizon spacetimes for arbitrary horizon radius and $z$.

\section{Energy-entropy relation for arbitrary $z$}\label{sec:results}
\subsection{Procedure}
For each solution, we fit the asymptotic data with the asymptotic power law expansion in~\cite{Basu:2016vyz} for asymptotically Lifshitz spacetimes. We use this fit to calculate the leading order deviation of $u\cdot\chi$ from its global Lifshitz behavior near asymptotic infinity.  From the deviation we can read off a length scale analogous to the Schwarzschild radius, $r_s$, which appears in the asymptotic expansion. $r_s$ is related to the mass per unit transverse length (since these are planar black hole solutions) of the black hole via~\cite{Basu:2016vyz}
\begin{eqnarray}
	M=\frac{(z+1)r_s}{8\pi Gl^2}.
\end{eqnarray}
Since we have set $l=1$ we can therefore state that for our solutions
\begin{equation}
	M=\frac{(z+1)r_s}{8\pi G} \label{eq:M}
\end{equation}
In~\cite{Herrero-Valea:2020fqa} the temperature dependence for universal horizons was determined as
\begin{equation}
	T=\frac{(a\cdot\chi)_{\textsc{uh}}}{4\pi}. \label{eq:T}
\end{equation} 
Finally, we will assume that the entropy of the universal horizon (again per unit transverse length) is given by
\begin{equation} \label{eq:S}
    S=\frac{r_{UH}} {4G}.
\end{equation}
Given these results, we can calculate the relation between $M,T$, and $S$ for all $z$ from our solutions and determine the energy-entropy relation.

\subsection{Results}
\begin{figure}[h]
 	\centering
 	\subfigure[]{\includegraphics[width=0.2\textwidth]{"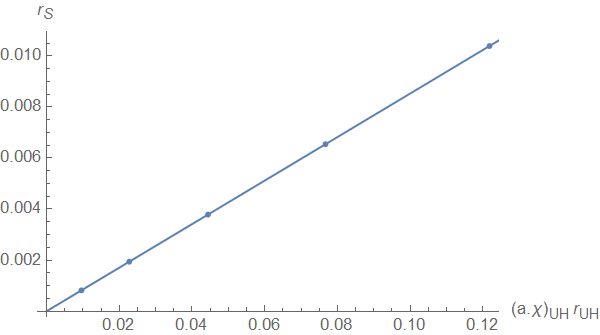"}} \subfigure[]{\includegraphics[width=0.2\textwidth]{"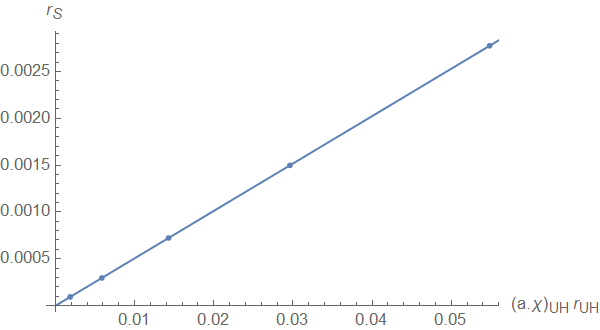"}} 	
  	\subfigure[]{\includegraphics[width=0.2\textwidth]{"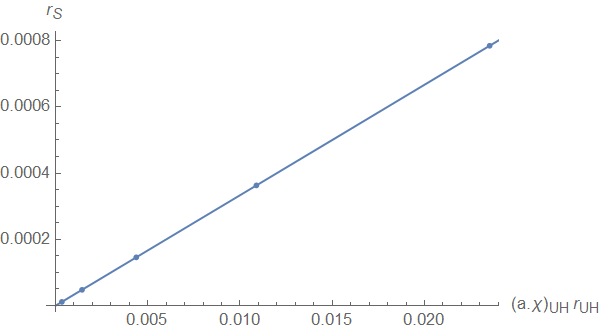"}} \subfigure[]{\includegraphics[width=0.2\textwidth]{"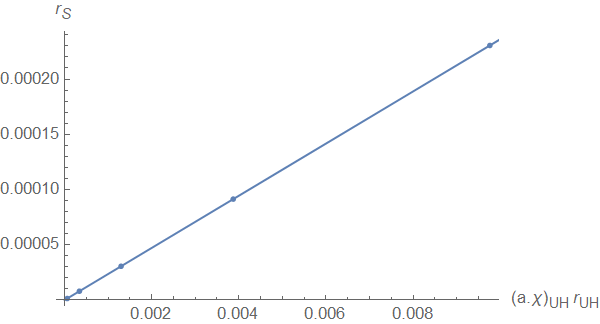"}} 
 	\subfigure[]{\includegraphics[width=0.2\textwidth]{"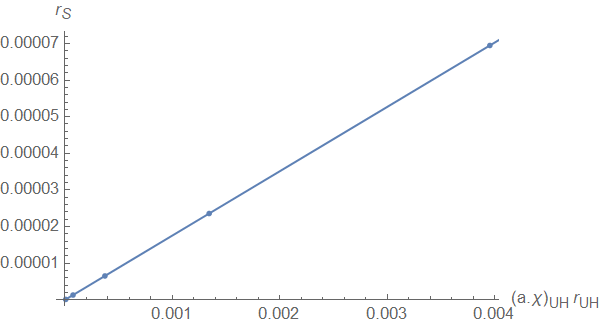"}} \subfigure[]{\includegraphics[width=0.2\textwidth]{"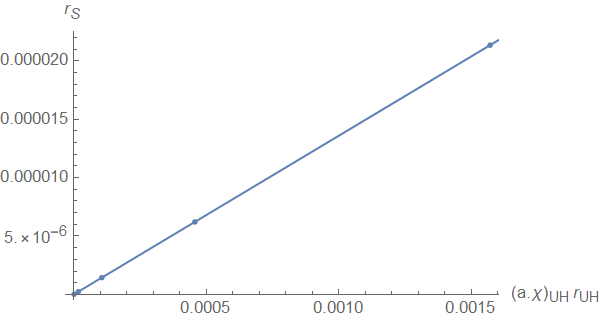"}}  	 	\subfigure[]{\includegraphics[width=0.2\textwidth]{"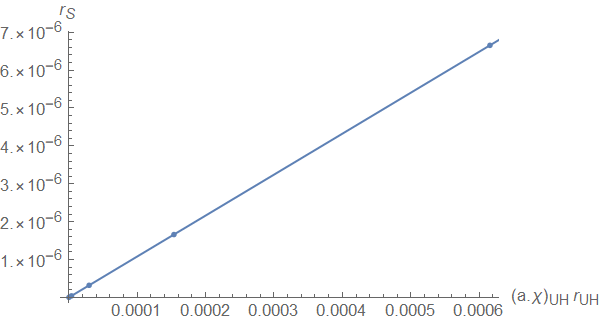"}} \subfigure[]{\includegraphics[width=0.2\textwidth]{"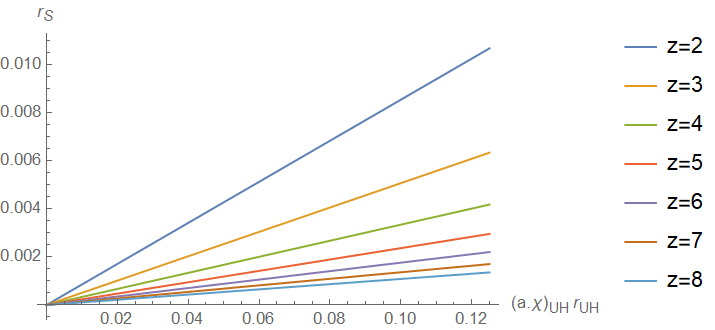"}} 
 	\caption{(a)-(g) Points show the obtained values of $r_s$ plotted against the universal horizon radius multiplied by $a\cdot\chi$ at the universal horizon, for each of five spacetimes found at each z=2...8. Lines show a linear fit to the data in each case. (h) The linear fit lines for each of z=2...8 plotted together.} 
 	\label{fig:firstlaws}
 \end{figure}

We first show that, as one would expect for a first law of black hole mechanics, the relationship between $r_s$ and $\ruh\times a\cdot\chi_{UH}$ is exactly linear, i.e. there exists a Smarr formula.  Here the subscript $UH$ means to evaluate at the universal horizon.  We plot $\ruh\times a\cdot\chi_{UH}$ vs. $r_s$ for five choices of $r_0$ and for all values of $z$ from 2 to 8 in Figure \ref{fig:firstlaws}.  The $z=2$ relation is equivalent to that in~\cite{Basu:2016vyz} while new results are for $z>2$.  For each value of $z$ it is hence clear that there exists a first law of the form $r_s \propto (a \cdot \chi) \delta \ruh$, and that the constant of proportionality varies with $z$.

\begin{figure}[h]
	\centering
	\includegraphics[width=0.4\textwidth]{"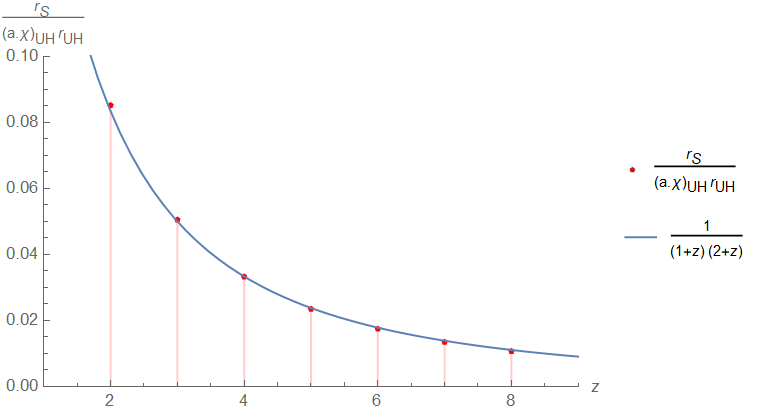"}
	\caption{Discrete points show the value of the ratio between $r_s$ and $\ruh\times(a\cdot\chi)_{\textsc{uh}}$. The curve shown is $\frac{1}{(z+1)(z+2)}$.}
	\label{fig:zdependence}
\end{figure}

 We now turn to the thermodynamical relationship based on a first law.  We expect a factor of $(z+1)$ in this proportionality due to the dependence of mass per length on $r_s$ in equation (\ref{eq:M}). Additional z dependence thus comes from the dynamics of \HLg~itself.  This dependence can be determined by fitting the $z$ dependence of the ratio of $r_s$ and $\ruh\times(a\cdot\chi)_{\textsc{uh}}$. As shown in Figure \ref{fig:zdependence}, this is extremely well approximated by $\frac{1}{(z+1)(z+2)}$. 

Using equations~\eqref{eq:M},\eqref{eq:T}, and~\eqref{eq:S} we therefore see that to leading order,
\begin{eqnarray}
	\frac{r_s}{\ruh(a\cdot\chi)_\textsc{uh}}&=&\frac{1}{(z+1)(z+2)}
\\	\frac{(z+1) r_s}{8\pi G}&=&\frac{2}{(z+2)}\frac{(a\cdot\chi)_\textsc{uh}}{4\pi}\frac{\ruh}{4G}
\\	 M&=&\frac{2}{z+2}T S
\end{eqnarray}
which is the main result of this work as this matches the energy-entropy relation found in \cite{Bertoldi:2009dt} for Lifshitz field theories. 

\begin{figure}[h]
	\centering
	\includegraphics[width=0.4\textwidth]{"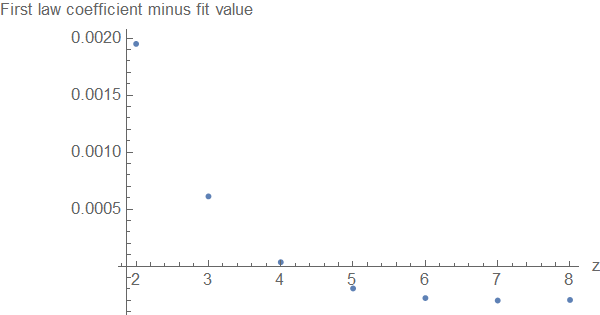"}
	\caption{Points show the difference between the value of the ratio between $r_s$ and $\ruh\times(a\cdot\chi)_{\textsc{uh}}$ and $\frac{1}{(z+1)(z+2)}$.}
	\label{fig:deviation}
\end{figure}

There \textit{is} a small deviation from this fit (plotted in Figure \ref{fig:deviation}), which does not appear random, and is therefore likely not an artifact of error in the numerically generated solutions. In principle there are subdominant contributions to the entropy for universal horizons, just as there are logarithmic corrections to black hole entropy in general relativity~\cite{Carlip:2000nv}. These corrections have, to our knowledge, not yet been calculated, although theoretically they may be able to via the Lifshitz extension of the Cardy formula~\cite{Gonzalez:2011nz} applied to universal horizons, just as the Cardy formula can be used to find the logarithmic corrections at Killing horizons. We leave finding the explanation of the deviation in terms of subdominant corrections or other physics for future work.

\section{Conclusion}\label{sec:implications}
Our results establish an energy-entropy relationship for universal horizons in 2+1 asymptotically Lifshitz spacetime of the form 
\begin{equation}
    E=\frac{2}{z+2}TS
\end{equation}
which is in agreement with that expected for Lifshitz quantum field theories in two dimensions. This extends previous results for $z=2$, concretely indicates there exists a thermodynamics for universal horizons with these asymptotics for any $z$, and provides evidence that the holographic dual is a Lifshitz field theory with matching dynamical exponent $z$.  While similar results have been found to hold for other relativistic gravitational theories, problems with relativistic bulk gravitational theories as holographic duals to Lifshitz field theories have been shown to exist (c.f. \cite{Griffin:2012qx},~\cite{Cheyne:2017bis}). Our result provides further evidence that the most natural dual of a Lifshitz field theory is \HLg~and that the bulk solutions appropriate for duals to thermal states are those with universal horizons at the corresponding temperature.

\begin{acknowledgements}
	This work was partially supported by the United States Department of Energy under award DE-SC0020220.
\end{acknowledgements}

\bibliographystyle{unsrt}

\begin{thebibliography}{99}

	\bibitem{Julian:1996}
	S R Julian et al
	J. Phys.: Condens. Matter \textbf{8} 9675 (1996)
	doi:10.1088/0953-8984/8/48/002
	
	%\cite{Stewart:2001zz}
    \bibitem{Stewart:2001zz}
    G.~R.~Stewart,
    %``Non-Fermi-liquid behavior in d- and f-electron metals,''
    Rev. Mod. Phys. \textbf{73}, 797-855 (2001)
    doi:10.1103/RevModPhys.73.797
    %105 citations counted in INSPIRE as of 01 Feb 2021
		
	\bibitem{Si:2008}
	P.~Gegenwart, Q.~Si, and F.~Steglich,
	%Quantum criticality in heavy-fermion metals.
	Nature Phys \textbf{4}, 186–197 (2008). 
	doi:10.1038/nphys892
	
	%\cite{Kachru:2008yh}
    \bibitem{Kachru:2008yh}
    S.~Kachru, X.~Liu and M.~Mulligan,
    %``Gravity duals of Lifshitz-like fixed points,''
    Phys. Rev. D \textbf{78}, 106005 (2008)
    doi:10.1103/PhysRevD.78.106005
    [arXiv:0808.1725 [hep-th]].
    %859 citations counted in INSPIRE as of 02 Feb 2021

	%\cite{Si:2011hh}
    \bibitem{Si:2011hh}
    Q.~Si and F.~Steglich,
    %``Heavy Fermions and Quantum Phase Transitions,''
    Science \textbf{329}, 1161 (2010)
    doi:10.1126/science.1191195
    [arXiv:1102.4896 [cond-mat.str-el]].
    %35 citations counted in INSPIRE as of 01 Feb 2021
	
	%\cite{Goldstein:2009cv}
	\bibitem{Goldstein:2009cv} 
	K.~Goldstein, S.~Kachru, S.~Prakash and S.~P.~Trivedi,
	%``Holography of Charged Dilaton Black Holes,''
	JHEP {\bf 1008}, 078 (2010)
	doi:10.1007/JHEP08(2010)078
	[arXiv:0911.3586 [hep-th]].
	%%CITATION = doi:10.1007/JHEP08(2010)078;%%
	%248 citations counted in INSPIRE as of 28 Jun 2017
	
	%\cite{Taylor:2008tg}
	\bibitem{Taylor:2008tg} 
	M.~Taylor,
	%``Non-relativistic holography,''
	arXiv:0812.0530 [hep-th].
	%%CITATION = ARXIV:0812.0530;%%
	%383 citations counted in INSPIRE as of 28 Jun 2017

	%\cite{Cheyne:2017bis}
	\bibitem{Cheyne:2017bis} 
	J.~Cheyne and D.~Mattingly,
	%``Constructing entanglement wedges for Lifshitz spacetimes with Lifshitz gravity,''
	Phys.\ Rev.\ D {\bf 97}, no. 6, 066024 (2018)
	doi:10.1103/PhysRevD.97.066024
	[arXiv:1707.05913 [gr-qc]].
	%%CITATION = doi:10.1103/PhysRevD.97.066024;%%
	%5 citations counted in INSPIRE as of 15 Jul 2019
	
	%\cite{Gentle:2015cfp}
	\bibitem{Gentle:2015cfp} 
	S.~A.~Gentle and C.~Keeler,
	%``On the reconstruction of Lifshitz spacetimes,''
	JHEP {\bf 1603}, 195 (2016)
	doi:10.1007/JHEP03(2016)195
	[arXiv:1512.04538 [hep-th]].
	%%CITATION = doi:10.1007/JHEP03(2016)195;%%
	%15 citations counted in INSPIRE as of 15 Jul 2019

	%\cite{AyonBeato:2009nh}
	\bibitem{AyonBeato:2009nh} 
	E.~Ayon-Beato, A.~Garbarz, G.~Giribet and M.~Hassaine,
	%``Lifshitz Black Hole in Three Dimensions,''
	Phys.\ Rev.\ D {\bf 80}, 104029 (2009)
	doi:10.1103/PhysRevD.80.104029
	[arXiv:0909.1347 [hep-th]].
	%%CITATION = doi:10.1103/PhysRevD.80.104029;%%
	%137 citations counted in INSPIRE as of 28 Jun 2017
	
	%\cite{Goya:2014eya}
	\bibitem{Goya:2014eya} 
	A.~F.~Goya,
	%``Anisotropic Scale Invariant Spacetimes and Black Holes in Zwei-Dreibein Gravity,''
	JHEP {\bf 1409}, 132 (2014)
	doi:10.1007/JHEP09(2014)132
	[arXiv:1406.4771 [hep-th]].
	%%CITATION = doi:10.1007/JHEP09(2014)132;%%
	%4 citations counted in INSPIRE as of 28 Jun 2017

	%\cite{Horava:2009vy}
	\bibitem{Horava:2009vy}
	P.~Horava and C.~M.~Melby-Thompson,
	%``Anisotropic Conformal Infinity,''
	Gen. Rel. Grav. \textbf{43}, 1391-1400 (2011)
	doi:10.1007/s10714-010-1117-y
	[arXiv:0909.3841 [hep-th]].
	%66 citations counted in INSPIRE as of 29 Jan 2021
		
	%\cite{Basu:2016vyz}
	\bibitem{Basu:2016vyz} 
	S.~Basu, J.~Bhattacharyya, D.~Mattingly and M.~Roberson,
	%``Asymptotically Lifshitz spacetimes with universal horizons in $(1 + 2)$ dimensions,''
	Phys.\ Rev.\ D {\bf 93}, no. 6, 064072 (2016)
	doi:10.1103/PhysRevD.93.064072
	[arXiv:1601.03274 [hep-th]].
	%%CITATION = doi:10.1103/PhysRevD.93.064072;%%
	%7 citations counted in INSPIRE as of 28 Jun 2017

	%\cite{Bertoldi:2009dt}
	\bibitem{Bertoldi:2009dt}
	G.~Bertoldi, B.~A.~Burrington and A.~W.~Peet,
	%``Thermodynamics of black branes in asymptotically Lifshitz spacetimes,''
	Phys. Rev. D \textbf{80}, 126004 (2009)
	doi:10.1103/PhysRevD.80.126004
	[arXiv:0907.4755 [hep-th]].
	%90 citations counted in INSPIRE as of 28 Jan 2021
		
	\bibitem{Horava:2009uw} 
	P.~Ho\v{r}ava,
	%``Quantum Gravity at a Lifshitz Point,''
	Phys.\ Rev.\ D {\bf 79}, 084008 (2009)
	doi:10.1103/PhysRevD.79.084008
	[arXiv:0901.3775 [hep-th]].
	%%CITATION = doi:10.1103/PhysRevD.79.084008;%%
	%1470 citations counted in INSPIRE as of 28 Jun 2017
	
	%\cite{Horava:2008ih}
	\bibitem{Horava:2008ih} 
	P.~Horava,
	%``Membranes at Quantum Criticality,''
	JHEP {\bf 0903}, 020 (2009)
	doi:10.1088/1126-6708/2009/03/020
	[arXiv:0812.4287 [hep-th]].
	%%CITATION = doi:10.1088/1126-6708/2009/03/020;%%
	%608 citations counted in INSPIRE as of 15 Jul 2019	
	
	%\cite{Sotiriou:2010wn}
	\bibitem{Sotiriou:2010wn} 
	T.~P.~Sotiriou,
	%``Ho\v{r}ava-Lifshitz gravity: a status report,''
	J.\ Phys.\ Conf.\ Ser.\  {\bf 283}, 012034 (2011)
	doi:10.1088/1742-6596/283/1/012034
	[arXiv:1010.3218 [hep-th]].
	%%CITATION = doi:10.1088/1742-6596/283/1/012034;%%
	%138 citations counted in INSPIRE as of 28 Jun 2017
	
	%\cite{Blas:2009qj}
	\bibitem{Blas:2009qj}
	D.~Blas, O.~Pujolas and S.~Sibiryakov,
	%``Consistent Extension of Horava Gravity,''
	Phys. Rev. Lett. \textbf{104}, 181302 (2010)
	doi:10.1103/PhysRevLett.104.181302
	[arXiv:0909.3525 [hep-th]].
	%481 citations counted in INSPIRE as of 29 Jan 2021
	
	%\cite{Jacobson:2010mx}
	\bibitem{Jacobson:2010mx} 
	T.~Jacobson,
	%``Extended Ho\v{r}ava gravity and Einstein-aether theory,''
	Phys.\ Rev.\ D {\bf 81}, 101502 (2010)
	Erratum: [Phys.\ Rev.\ D {\bf 82}, 129901 (2010)]
	doi:10.1103/PhysRevD.82.129901, 10.1103/PhysRevD.81.101502
	[arXiv:1001.4823 [hep-th]].
	%%CITATION = doi:10.1103/PhysRevD.82.129901, 10.1103/PhysRevD.81.101502;%%
	%134 citations counted in INSPIRE as of 28 Jun 2017
	
	%\cite{Jacobson:2000xp}
	\bibitem{Jacobson:2000xp} 
	T.~Jacobson and D.~Mattingly,
	%``Gravity with a dynamical preferred frame,''
	Phys.\ Rev.\ D {\bf 64}, 024028 (2001)
	doi:10.1103/PhysRevD.64.024028
	[gr-qc/0007031].
	%%CITATION = doi:10.1103/PhysRevD.64.024028;%%
	%470 citations counted in INSPIRE as of 28 Jun 2017

	%\cite{Pacilio:2017emh}
    \bibitem{Pacilio:2017emh}
    C.~Pacilio and S.~Liberati,
    %``Improved derivation of the Smarr formula for Lorentz-breaking gravity,''
    Phys. Rev. D \textbf{95}, no.12, 124010 (2017)
    doi:10.1103/PhysRevD.95.124010
    [arXiv:1701.04992 [gr-qc]].
    %13 citations counted in INSPIRE as of 02 Feb 2021

	%\cite{Liberati:2017vse}
	\bibitem{Liberati:2017vse}
	C.~Pacilio and S.~Liberati,
	%``First law of black holes with a universal horizon,''
	Phys. Rev. D \textbf{96}, no.10, 104060 (2017)
	doi:10.1103/PhysRevD.96.104060
	[arXiv:1709.05802 [gr-qc]].
	%9 citations counted in INSPIRE as of 27 Jan 2021
	
	%\cite{Berglund:2012bu}
	\bibitem{Berglund:2012bu} 
	P.~Berglund, J.~Bhattacharyya and D.~Mattingly,
	%``Mechanics of universal horizons,''
	Phys.\ Rev.\ D {\bf 85}, 124019 (2012)
	doi:10.1103/PhysRevD.85.124019
	[arXiv:1202.4497 [hep-th]].
	%%CITATION = doi:10.1103/PhysRevD.85.124019;%%
	%71 citations counted in INSPIRE as of 16 Jul 2019
	
	%\cite{Berglund:2012fk}
	\bibitem{Berglund:2012fk}
	P.~Berglund, J.~Bhattacharyya and D.~Mattingly,
	%``Towards Thermodynamics of Universal Horizons in Einstein-\ae{}ther Theory,''
	Phys. Rev. Lett. \textbf{110}, no.7, 071301 (2013)
	doi:10.1103/PhysRevLett.110.071301
	[arXiv:1210.4940 [hep-th]].
	%69 citations counted in INSPIRE as of 28 Jan 2021

	%\cite{Bhattacharyya:2014kta}
	\bibitem{Bhattacharyya:2014kta} 
	J.~Bhattacharyya and D.~Mattingly,
	%``Universal horizons in maximally symmetric spaces,''
	Int.\ J.\ Mod.\ Phys.\ D {\bf 23}, no. 13, 1443005 (2014)
	doi:10.1142/S0218271814430056
	[arXiv:1408.6479 [hep-th]].
	%%CITATION = doi:10.1142/S0218271814430056;%%
	%18 citations counted in INSPIRE as of 28 Jun 2017

	%\cite{Griffin:2012qx}
	\bibitem{Griffin:2012qx} 
	T.~Griffin, P.~Hořava and C.~M.~Melby-Thompson,
	%``Lifshitz Gravity for Lifshitz Holography,''
	Phys.\ Rev.\ Lett.\  {\bf 110}, no. 8, 081602 (2013)
	doi:10.1103/PhysRevLett.110.081602
	[arXiv:1211.4872 [hep-th]].
	%%CITATION = doi:10.1103/PhysRevLett.110.081602;%%
	%78 citations counted in INSPIRE as of 28 Jun 2017

    %\cite{Blas:2011ni}
    \bibitem{Blas:2011ni}
    D.~Blas and S.~Sibiryakov,
    %``Horava gravity versus thermodynamics: The Black hole case,''
    Phys. Rev. D \textbf{84}, 124043 (2011)
    doi:10.1103/PhysRevD.84.124043
    [arXiv:1110.2195 [hep-th]].
    %116 citations counted in INSPIRE as of 02 Feb 2021	
    %\cite{Foster:2005ec}
    \bibitem{Foster:2005ec}
    B.~Z.~Foster,
    %``Metric redefinitions in Einstein-Aether theory,''
    Phys. Rev. D \textbf{72}, 044017 (2005)
    doi:10.1103/PhysRevD.72.044017
    [arXiv:gr-qc/0502066 [gr-qc]].
    %42 citations counted in INSPIRE as of 02 Feb 2021
	
    %\cite{Herrero-Valea:2020fqa}
    \bibitem{Herrero-Valea:2020fqa}
    M.~Herrero-Valea, S.~Liberati and R.~Santos-Garcia,
    %``Hawking Radiation from Universal Horizons,''
    [arXiv:2101.00028 [gr-qc]].
    %0 citations counted in INSPIRE as of 02 Feb 2021

	%\cite{Carlip:2000nv}
    \bibitem{Carlip:2000nv}
    S.~Carlip,
    %``Logarithmic corrections to black hole entropy from the Cardy formula,''
    Class. Quant. Grav. \textbf{17}, 4175-4186 (2000)
    doi:10.1088/0264-9381/17/20/302
    [arXiv:gr-qc/0005017 [gr-qc]].
    %324 citations counted in INSPIRE as of 03 Feb 2021

    %\cite{Gonzalez:2011nz}
    \bibitem{Gonzalez:2011nz}
    H.~A.~Gonzalez, D.~Tempo and R.~Troncoso,
    %``Field theories with anisotropic scaling in 2D, solitons and the microscopic entropy of asymptotically Lifshitz black holes,''
    JHEP \textbf{11}, 066 (2011)
    doi:10.1007/JHEP11(2011)066
    [arXiv:1107.3647 [hep-th]].
    %60 citations counted in INSPIRE as of 03 Feb 2021

	
\end{thebibliography}

\end{document}